# Topological time series analysis of a string experiment and its synchronized model

Nicholas B. Tufillaro
*Center for Nonlinear Studies and T13, MS-B258*
*Los Alamos National Laboratory, Los Alamos, NM 87545*

Peter Wyckoff
*Department of Chemical Engineering, MIT 16-311, Cambridge, MA 02139*

Reggie Brown
*Institute for Nonlinear Science, University of California, San Diego, Mail Code 0402, La Jolla, CA 92093-0402*

Thomas Schreiber
*Physics Department, University of Wuppertal, D-42097 Wuppertal, Germany*

Timothy Molteno
*Department of Physics, University of Otago, Dunedin, New Zealand*

We identify the template organizing the chaotic dynamics of a vibrating string and show how to estimate topological parameter values directly from an experimental time series. We call these data processing techniques "topological time series analysis." The chaotic motions of a vibrating string are shown to be governed by a dissipative horseshoe map whose periodic orbit spectrum is, at the topological level, well described by a unimodal map of the interval.

In addition, close topological agreement is demonstrated between the experimental data and a synchronized model. The empirical model of the experimental data allows us to create "synthetic data." Such synthetic data may enable a finer characterization of the system than that determined from the seed data alone.

## I. INTRODUCTION

Our studies of the nonlinear and chaotic vibrations of a string are motivated by two intertwining concerns. First, we want to know strings. Or at least what we can know of strings via experiments showing chaotic motions in a vibrating wire. Second, given an experimental times series from a low dimensional chaotic process, we would like to provide a quantitative characterization of the chaotic attractor—or at least those properties of the attractor which are possibly shared by a large class of similar chaotic systems. Such a characterization should be robust under the effects of noise and smooth parameter changes. Such a characterization is necessarily topological.

Interest in a string's motion has a long and illustrious history. In 1883 Kirchoff derived an inherently nonlinear differential equation which properly took into account the simultaneous longitudinal and transverse displacements of a physical string [1]. In 1968 Narashima rederived a scaled version of Kirchoff's equation [2]. Miles in 1984, beginning with Narashima's equation, derived and analyzed a four dimensional model of the averaged oscillations of a string and discovered a Hopf bifurcation [3]. In 1989, after careful numerical studies, Johnson and Bajaj detected the presence of chaotic orbits in this same model [4]. Guided by these theoretical insights, experimental detection of these chaotic oscillations soon followed [5,6]. More complete historical references are found in several recent Ph. D. studies on the nonlinear vibrations of a string [7–9].

Of particular interest to our experimental studies are the theoretical examinations of low dimensional nonlinear string models by Miles [3], Bajaj and Johnson [10], Tufillaro [11], and O'Reilly [6,12]; and the experimental reports of chaotic vibrations in a string by Molteno and Tufillaro [5], Molteno [9], and O'Reilly and Holmes [6]. As remarked by most of these authors, chaotic motions in a string are not easy to observe. They exist only in a small parameter range and occur not at the forcing frequency, but rather in the slow oscillations of the amplitude envelope. These amplitude modulations are usually only a fraction of the overall transverse displacement. These facts may help to account for the rather late discovery of chaotic vibrations in such a common system.

As recently advocated by several authors [13,14], the quantitative topological characterization of a low dimensional chaotic invariant set should proceed in two steps. First, the strange set needs to be assigned a good symbolic encoding. We address this problem by identifying the template organizing the stretching, twisting, and folding motion of the attractor in phase space [16]. Second, rules need to be developed specifying the presence or absence of subsets of symbol sequences when parameters are changed. The second problem is sometimes called pruning [15]. We present evidence that the "pruning problem" can be solved with predictions calculated by unimodal map theory—at least in the parameter regime considered and to within the resolution of the experimental measurements. Two types of evidence are used in support of these findings. The "symbol plane" is reconstructed for an experimental chaotic trajectory and it reveals that the system is well approximated by a vertical pruning front with no steps. Additionally, periodic orbits are extracted by the method of close recurrence [17,18] and their spectrum is ordered by unimodal theory. Since the data appears to be well described by unimodal theory, we also estimate the value of the kneading sequence, the "topological parameter" which fixes the spectrum of periodic orbits.

In addition, we construct an empirical global model of the dynamics that produced the time series by determining a vector which best fits the data [19]. The method used to fit the vector field is based on the minimum description length (MDL) criterion of Rissanen [20]. Two distinct methods are used in an attempt to address the question of how closely the empirical model fits the data. First, we show that we can synchronize the model to the data. Synchronizing these models to the data set provides the first piece of evidence that the empirical model is "close" to the flow observed experimentally [27]. Second, we show that the experimental data and a time series from the model share the same template and are close in terms of topological parameters.

This paper is organized as follows. Section II describes the string experiment and examines data sets taken from a parameter regime which leads to the onset of a crisis. Section III briefly describes how to construct a global empirical model from a data set. The Lyapunov spectra of the model and the data set are also calculated and it is suggested that calculating such system characteristics indirectly from "synthetic data" is a desirable and sensible procedure. Section IV describes how periodic orbit spectra are extracted by the method of close recurrence for data sets at two distinct parameter values. It also describes how braid invariants are used to identify the template organizing the periodic and chaotic orbits of the strange attractor. A topological analysis of data generated from both the experiment and the empirical model is then presented. We call these data processing techniques "topological time series analysis," since they seek to ascertain topological form and topological parameter values directly from an experimental or simulated time series. Some concluding remarks are offered in Section V.

Lastly, we would like to remark that our experimental results show good qualitative agreement with the numerical simulations on nonlinear and chaotic vibrations in strings by Bajaj and Johnson [10]. This correspondence is discussed more fully in Ref. [9].

## II. EXPERIMENTAL SETUP AND DATA SET DESCRIPTION

A schematic of the experimental apparatus used to excite and detect forced vibrations in a thin wire is shown in Fig. 1.

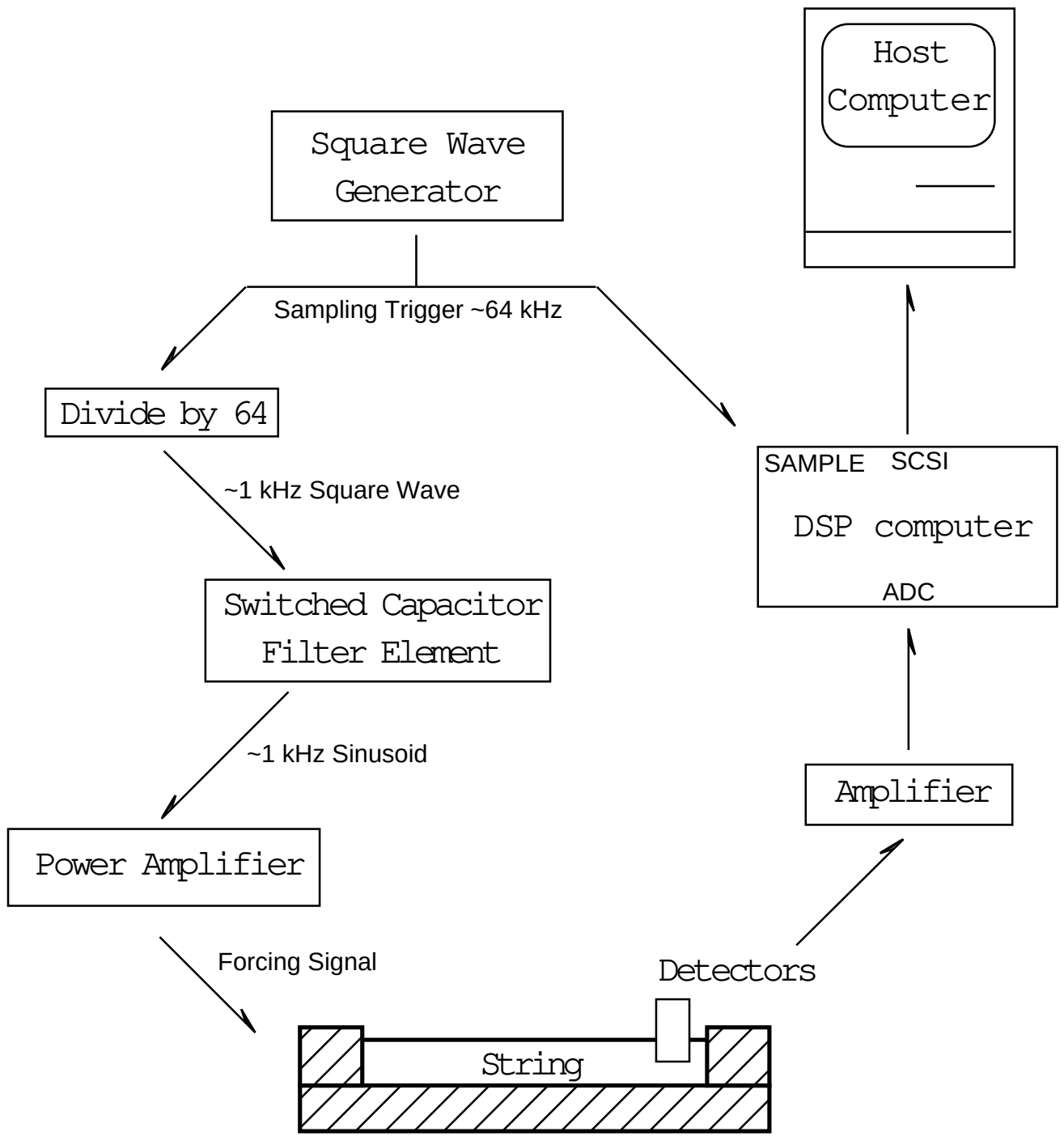

FIG. 1. Schematic of experimental apparatus used to excite and detect forced vibrations in a thin wire.

This apparatus is a considerably improved version of that described in Ref. [5]. A non-magnetic thoriated tungsten wire, 0.15 mm in diameter, is fastened in a rigid mount. The wire sits in a permanent magnetic field created with rare earth ceramic magnets. An alternating current is passed through the wire to excite vibrations. Optoelectronic motion detectors are used to measure the transverse wire displacement simultaneously in two orthogonal directions. The detectors consist of an infrared photodiode coupled to a matched phototransistor positioned to detect the partial occlusion of the beam by the wire.

Both the forcing current and detected signal are handled with a custom built controller/detector system designed around an Analog Devices ADSP 2105 Digital Signal Processor (DSP). It is a 16-bit device with a pipelined Harvard Architecture capable of performing 30 million operations per second. Direct Digital Synthesis (DDS) is used to generate the signal for forcing the wire. This technique is used because fine control over the forcing frequency is required to characterize the constant amplitude response of the wire. This digital control and detection scheme also allows the development of real-time nonlinear signal processing and identification tools such as variable (in phase and amplitude) Poincaré sections,

Fast Fourier Transforms, and embedding plots with variable delays. All these real-time nonlinear signal processing tools greatly aid in the detection, identification, and optimization of the measurements reported here.

Typical parameter values for the wire and apparatus are: wire length, 0.07 m; mass per unit length, $3.39 \times 10^{-4}$ kg m$^{-1}$; density $2.1 \times 10^{4}$ kg m$^{-3}$; Young's modulus, $197778 \times 10^{6}$ N m$^{-1}$; magnetic field strength, 0.2 Tesla; and a free frequency of vibration in the neighborhood of 1350 Hz. The damping of the system is measured from the decay rate of free vibrations and can be modified by the application of a silicone coating to the wire [6].

The transverse amplitude displacement of the wire is usually sampled once each forcing cycle at a fixed phase ($f_d \approx 1300$ Hz). The response frequency of the amplitude modulation is about 10 Hz. Thus, we have about 130 samples per cycle—the system is over sampled. Such over sampling is necessary, however, in order to calculate accurately some of the braid (extracted periodic orbit) invariants described in the Section IV. Since the interesting dynamics occurs so slowly, it is possible to save large data sets directly to disk; data sets of $10^5$ measurements are easy to obtain. The amplitude displacement measurements are usually low-noise (less than 1% of the RMS amplitude) and make good use of the wide dynamic range available. However, two measurement problems are present. First the system is subject to a parametric drift (typically manifested as a 1% variation of the RMS amplitude over 200 seconds) which we believe is due to a temperature instability in the drive system. Second, there are some systematic measurement errors due to phase jitter in the digital detecting electronics. Overall, though, the measurements are of high quality: large, low noise, data sets making good use of the wide dynamic range made available from the 16-bit digitizers.

After being recorded the raw data is "dynamically" cleaned by the nonlinear noise reduction technique described in Ref. [24]. The scalar data is embedded using eleven dimensional delay coordinates. For each data point a small neighborhood is formed in which the two dimensional submanifold spanned by the attractor is approximated by a linear subspace. Projections onto this subspace yield an improved time series. The procedure is iterated six times and curvature corrections are applied.

In addition, dynamical cleaning provides an estimate of the noise level. For the data we have examined, the amplitude of the correction, which can be taken as an estimate for the amplitude of the noise in the data, is 55 units (0.44%). We also estimate the amount of noise in the data independently by the method given in [25]. The functional form of the increase of the correlation integral with embedding dimension is known for Gaussian measurement noise, so that the width of the Gaussian distribution can be determined by a simple function fit. We found that indeed the errors are compatible with a Gaussian distribution of width 65 units (0.5%). After nonlinear noise reduction the errors are no longer expected to be Gaussian but the remaining noise level is found to be at most 25 units (0.2%).

As the first step toward extracting a template from the time series, a three dimensional embedding of a chaotic trajectory is created out of the scalar amplitude measurements made by a single detector. This is accomplished via a time delayed embedding of the original scalar data set [21]. The offset for the delay is determined by the mutual information criterion [22]. Values for the offset range from 25 to 39 for the data sets examined, or, not unexpectedly, about one quarter of the samples per chaotic cycle. A false nearest neighbor test also indicates that the time series is embeddable in three dimensions [21]. Figs. 2 shows three trajectories that are realized after a torus-doubling route to chaos [5]. Figs. 2(a) and 2(b) show chaotic attractors, and Fig. 2(c) shows a trajectory after the attractor has been extinguished by a crisis with a remote fixed point.

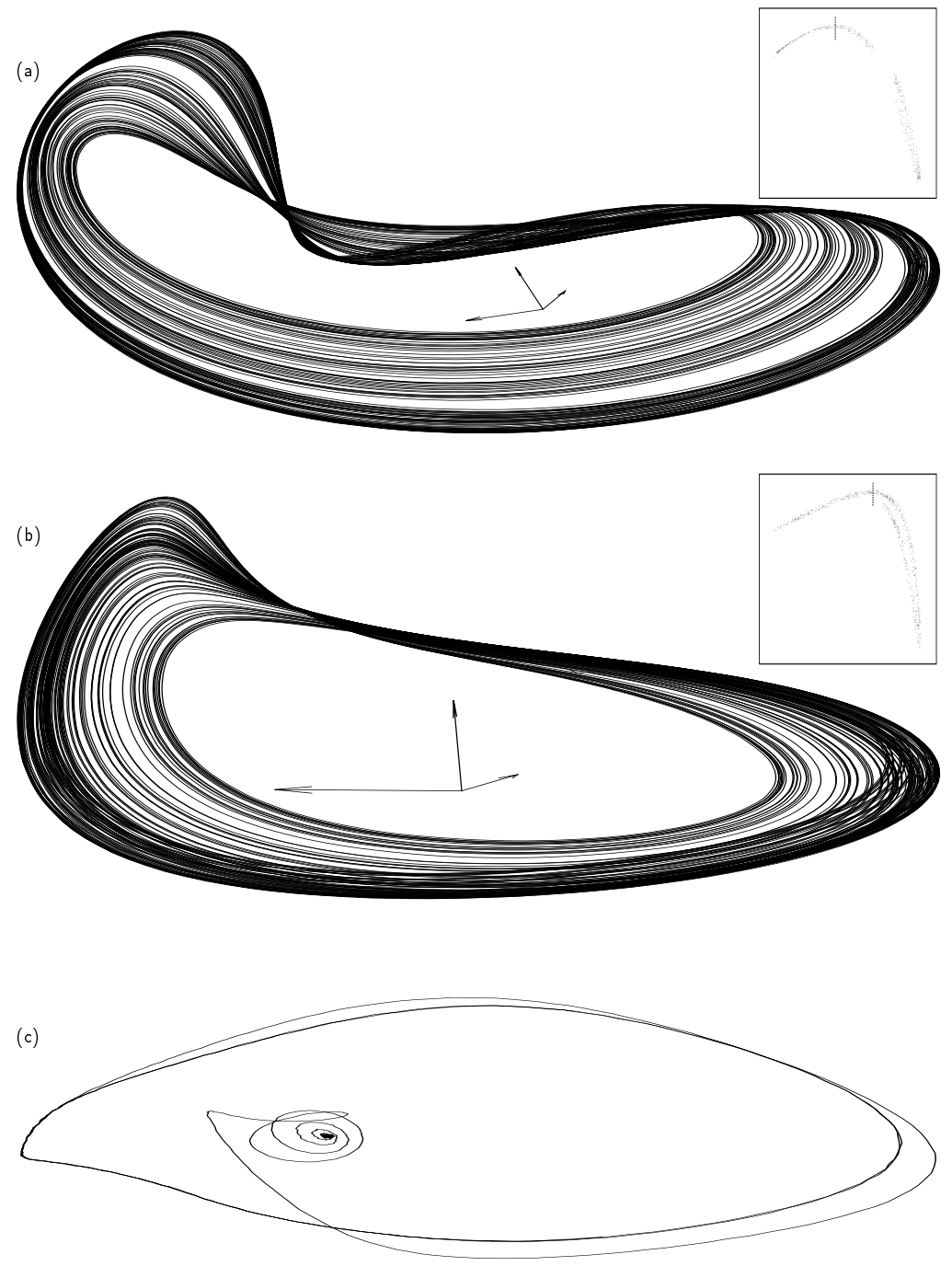

FIG. 2. Three dimensional embedded time series from the string experiment at three different parameter values: (a) chaotic data set (a); chaotic data set (b); (c) crisis with a remote fixed point. The insets show the low dimensional character of the next return maps constructed from a surface of section of the embedded attractors.

These experimental results are in good qualitative agreement with the numerical simulations of Bajaj and Johnson [10]. Fractal dimension calculations of these chaotic time series indicate a dimension of 2.1 [23]. Perhaps the strongest evidence, though, of the low dimensionality of these data sets is obtained by constructing a first return map of a two dimensional Poincaré surface of section: these are shown in the insets of Figs. 2(a) and 2(b).

## III. A SYNCHRONIZED EMPIRICAL MODEL

Our goal in this section is to create an empirical (global) model of the time series data. We do this by generating a vector field for a three dimensional system of ordinary differential equations whose dynamics "mimics" the dynamics of the experimental time series. The ability to "learn the evolution rules" only from *a priori* information is a powerful technique which opens the door to further applications such as short-term prediction and control. Also, as suggested here, the empirical model can be used to generate large, low noise, "synthetic" data sets. Such a synthetic data set, or the analytic knowledge supplied by the model, often permits a more refined calculation of a system's characteristics, such as its Lyapunov spectrum. However, before putting too much trust in the model, we must address the question of how faithfully the empirical model represents the dynamics that generated the data. We approach this question by two distinct methods. In this section we show that the model can be synchronized to the data [26]. In the next section we compare topological properties extracted from the experimental time series to the properties extracted from a time series generated by the model.

The method used to generate an empirical model directly from a data set is described in detail in a series of papers by Brown and co-workers [19]. The method assumes that the dynamics that produced the data vectors, $\mathbf{y}$, can be written as a set of ordinary differential equations

$$\frac{d\mathbf{y}}{dt} = \mathbf{F}(\mathbf{y}).$$

The vector field, $\mathbf{F}$, is written as a expansion in terms of polynomials that are orthonormal on the attractor represented by the experimental data

$$\mathbf{F}(\mathbf{y}) = \sum_{\mathbf{I}=\mathbf{0}}^{\mathbf{N_P}} \mathbf{p}^{(\mathbf{I})} \pi^{(\mathbf{I})}(\mathbf{y}).$$

These polynomials, $\pi^{(\mathbf{I})}$ form a basis set in which to write a model. The expansion coefficients, $\mathbf{p}^{(\mathbf{I})}$ are trained by modeling the system's dynamics as an implicit Adams integration scheme

$$\mathbf{y}(t+\tau) = \mathbf{y}(t) + \tau \sum_{j=0}^{M} a_j^{(M)} \mathbf{F}\big[\mathbf{y}(t-(j-1)\tau)\big],$$

where the $a_j^{(M)}$'s are the implicit Adams coefficients of order $M$.

In the traditional numerical integration scheme the coefficients, $\mathbf{p}^{(\mathbf{I})}$'s, are viewed as being fixed. In contrast, in the modeling context, we view these coefficients as being variable, their values being chosen to ensure a good correspondence between the experimental data and the empirical model. The model is trained by minimizing an MDL-type functional [20] which includes both a least squares minimization term and terms associated with the size and order of the vector field used to fit the data. The advantage of using an MDL-type functional is that an *optimal* model, from the class of polynomial models, can be found. Heuristically, the parameterization chosen is optimal in that it best fits the data with fewest number of terms. Higher order models may provide more accuracy, but the cost of the additional terms in the MDL functional outweighs the profit gained by the increase in accuracy.

To illustrate this modeling technique we consider the data set shown in Fig. 2(b). It consists of $128 \times 10^3$ scalar 16-bit measurements of the vertical transverse string displacement. The forcing frequency in this particular example is 1.384 kHz and the characteristic response frequency of the amplitude modulation is about 9 Hz. The average mutual information and false near neighbor methods indicates that the optimal time delay and embedding dimension are $T = 39$ and $d = 3$ [21]. The model is trained by using a subsection of $10^4$ points of the entire data set. More details about the modeling of this particular data set can be found in Ref. [27].

As suggested in Ref. [19] after the model is constructed we then use the experimental time series to *drive* the empirical model. We find that it the experimental time series is used to drive the model via dissipative coupling then the model synchronizes to the experimental time series. A detailed study of the quality of synchronization between the model and the data in the presence of additive noise and drift in the dynamics of the driving signal is presented in Ref. [27]. Synchronization between model and data provides the first piece of evidence that the empirical model and the experimental system are, in some respects, close.

We have also calculated the spectrum of Lyapunov exponents for the model as well as the data. Since it is the easier of the two calculations we first calculate the Lyapunov spectrum of the empirical model using a modified QR method [28–31]. We only report the two nonzero exponents of the three dimensional model. The results are shown in Figs. 3(a) and 3(b) where $K$ represents the total number of Jacobians used for the calculations (the total evolution time is $K\tau$ where $\tau = 0.05$ is the normalized time between the data points). The figures indicate that for $K > 5000$ the values of $\lambda_1$ and $\lambda_3$ are essentially constant. We have used a small ensemble of ten different initial conditions taken from disparate regions of the attractor for our calculations. The error bars in the figures indicate the maximum and minimum values of the Lyapunov exponents calculated from this ensemble. The ensemble is small, however the error bars indicate that the variance of the calculated values of the Lyapunov exponents over the initial conditions is also small. Furthermore, if the initial conditions for our calculations are within the basin of attraction of the attractor then the Lyapunov exponents are independent of the initial conditions in the $K\tau \to \infty$ limit [28,30].

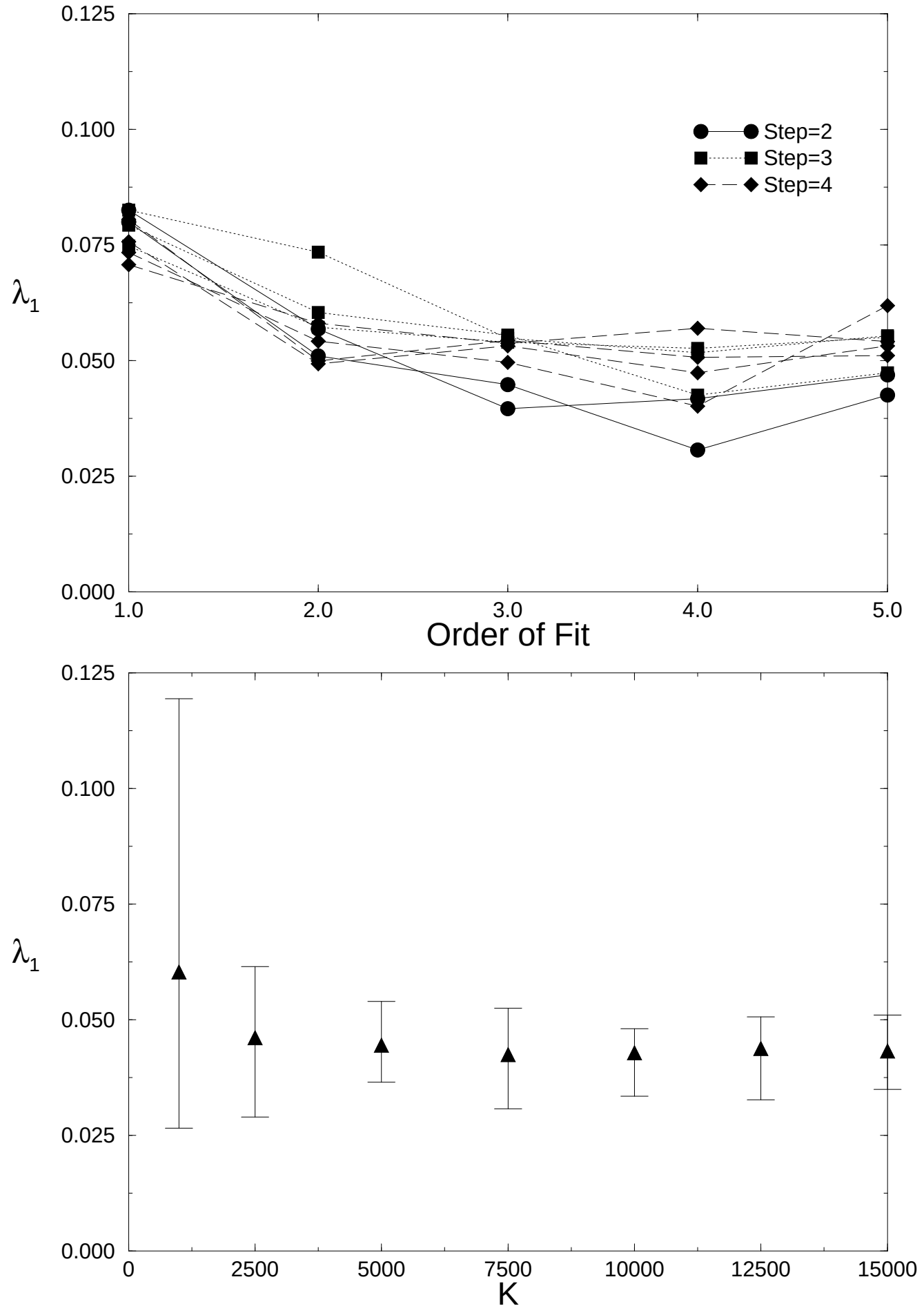

FIG. 3. Lyapunov spectrum for empirical model.

The relative independence of the calculated values of the Lyapunov exponents for $K > 5000$ and different initial conditions provide circumstantial evidence that our calculations have reached the asymptotic regime. We report the following values for the Lyapunov exponents of the empirical model: $\lambda_1 = 4.31 \times 10^{-2}$, and $\lambda_3 = -0.576$. These values are obtained by averaging over the ten initial conditions for $K = 15,000$.

Calculating the full spectrum of Lyapunov exponents from an experimental data set is a notoriously difficult procedure. This is particularly true with regard to the negative exponents. In general, the best that one can hope for is to calculate a spectrum that is consistent with itself and whatever additional facts are known about the dynamics. The method we use to calculate the Lyapunov exponents from the experimental data is a minor variation of the one reported in Ref. [32]. This method uses polynomials to map local neighborhoods on the attractor into their time evolved images.

The first variation we employ involves using the modified QR method reported in Refs. [30,31] instead of the one reported in Ref. [32]. The second modification is more complex and involves the data used to form the attractor. This second modification is required because the original data set is highly over sampled and needs to be "thinned out" before a reliable estimate of the spectrum can be obtained. The initial calculations use every other data *vector* to form the attractor. Next, every third data *vector* is used to form the attractor. Finally, every fourth data *vector* is used to form the attractor.

The second modification also results in a change in the evolution time (we call it the step) associated with the polynomial maps from one local neighborhood to its image. We find that as the step increases the influence of experimental noise is reduced and consistent values are obtained for the Lyapunov exponents. In addition, the number of possible initial conditions increases. When we use every second data vector we have two different initial conditions, $\mathbf{y}(1)$ and $\mathbf{y}(2)$. Similarly when we use every third data vector we have three different initial conditions, etc. We use the different initial conditions as another consistency check for our calculations. For each case the number of Jacobians used in the calculation is $K \simeq N$.

In all cases the total number of vectors used to form the attractor is $N \simeq 10,000$. Therefore, the size of the local neighborhoods used for the polynomial fitting is essentially the same for all tests. This insures consistency over our various tests. The value of $N \simeq 10,000$ is chosen on the *ad hoc* belief that larger values of $N$ will result in local neighborhoods that are so small that their dynamics will be unduly influenced by noise in the data [33,34].

The results of these calculations are presented in Figs. 4(a) and 4(b). When the order of the fit is greater than two the calculated values of the Lyapunov exponents become independent of the order of the fit. The only exception occurs for $\lambda_3$ and the attractor obtained by using every other data vector, step = 2. Since these values are associated with the negative Lyapunov exponent, and are well outside the range of the other results, we will ignore them. As one would expect, there is some spread in the calculated values of the Lyapunov exponents depending on the initial condition used and the different evolution times (step = 2, 3, or 4). In general however, the spread is small for both the positive *and* the negative Lyapunov exponent. As a final matter we note that the total evolution time for our calculation is $2K\tau$ when we use every other data point, $3K\tau$ when we use every third data point, etc. The independence of our results over step = 2, 3, and 4 indicates that $K \simeq 10,000$ corresponds to the time asymptotic regime. The consistency of our results over changes in initial conditions, order of the fitting, and the evolution time leads us to conclude that the Lyapunov exponents calculated from the data set using this method are close to the true exponents.

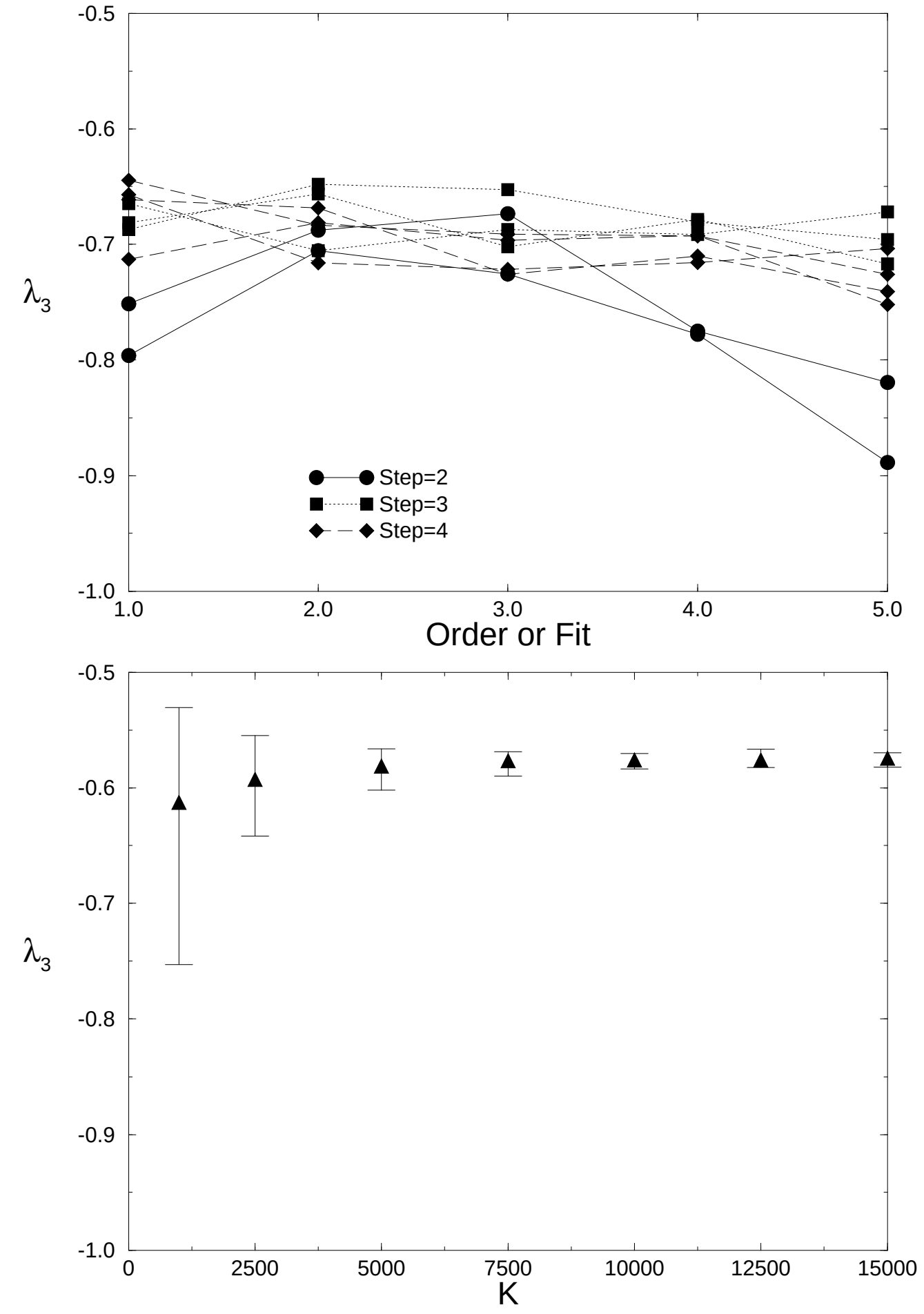

FIG. 4. Lyapunov spectrum from experimental data.

To obtain numerical values of the Lyapunov exponents we first averaged the calculated values of $\lambda$ over the different initial conditions and then averaged that value over the order of the fit for all fits greater than 2. We find that $\lambda_1 = 4.97 \times 10^{-2}$ and $\lambda_3 = -0.702$. These values differ from those reported for the model by approximately 15% for the positive exponent and 20% for the negative exponent.

## IV. TOPOLOGICAL TIME SERIES ANALYSIS

### A. Template identification

As the first step in a "topological time series analysis," we attempt to identify a "template" [16,35,36] organizing the periodic orbit structure of the flow. A template is nothing more or less than a cartoon of the stretching and folding structure of the flow written in a canonical form [38].

Before constructing the template, we note that the single-hump form of the first return map shown in the insets of Figs. 2(a) and (b) suggests that a "once-fold" or "horseshoe" like map [15] is organizing the global dynamics. Hence, one does not need the template to ascertain this topological information. However, the template contains an additional piece of topological information, namely, the global torsion, which can not be found by simply examining the first return map.

To ascertain the form of the template we proceed in a straightforward way. First, we examine the data in the three dimensional embedding by rotating it, and viewing it from several viewpoints with the aid of a graphics program we have written [37]. This graphical visualization suggests that the sheeted structure shown in Fig. 5(a) properly portrays the stretching, twisting, and folding of phase space which is organizing the orbit structure.

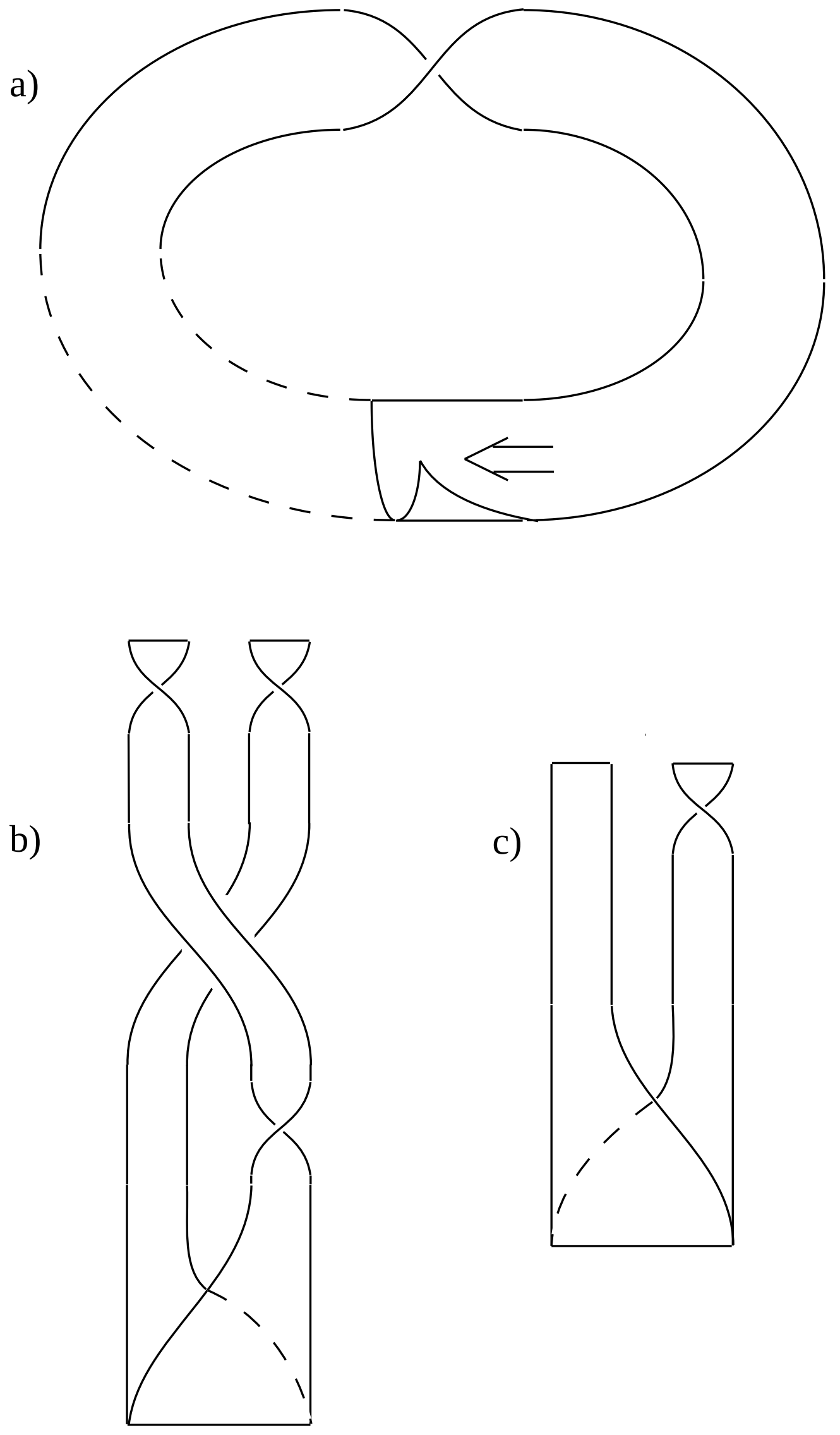

FIG. 5. Transformations on a horseshoe template: (a) Schematic of sheeted structure in the experimental data; (b) The template obtained from (a); (c) Horseshoe template equivalent to (a).

The (negative) half-twist at the top of Fig. 5 is followed by a small (positive) fold shown at the bottom (compare with Fig. 5(b)). This small fold is responsible for the stretching (causing sensitive dependence) and folding (causing the periodic and aperiodic recurrence) that produces the chaos. Both of these features of the flow are themselves organized by basic topological properties of the flow, namely, the flow's fixed points and their homoclinic and heteroclinic connections.

To arrive at the canonical form of the template we separate the sheet shown in Fig. 5(a) into two branches by splitting along the trajectory of the fold point (in the language of Ref. [15], we are creating a symbolic partition by considering the outermost "primary tangency"). Next we pull the small fold all the way to the bottom (thus going from a pruned to an unpruned system). The (negative) half twist in Fig. 5(a) results in a half twist in each branch, with both branches crossing as shown in the top of Fig. 5(b). The bottom of Fig. 5(b) shows how the small fold results in a horseshoe template in standard form [38]. Now we push all the branch twists in Fig. 5(b) to the top of the diagram and note that the twists in the left branch cancel, so that we arrive at the template shown in Fig. 5(c). We note that this is itself a horseshoe template with a global torsion of zero. Our construction thus shows that a standard horseshoe template with a global negative half twist is again a horseshoe template.

In Subsection C we will extract periodic orbits from chaotic time series and discuss how each orbit is given a symbolic itinerary. Given these periodic orbits, we then attempt to verify (or falsify) the conjecture that a horseshoe template with zero global torsion is organizing the dynamics. Specifically, we check two types of invariants: linking numbers of periodic orbits (a knot invariant), and the exponent sum (an invariant on *positive* braids) calculated from the "natural" braid associated to each periodic orbit [38–40]. As noted by Hall [41], the *exponent sum is a complete braid invariant for all horseshoe braids up to period eight*—thus, this single invariant tells us the symbolic name (up to saddle-node pairs) of the low period orbits in a horseshoe without the need for an empirical symbolic partition. This is an effective procedure on orbits all of whose crossings are clearly resolved. (These are all the orbits examined up to period eleven in the model, but only orbits up to period six in the experimental data.) We have found no discrepancies between the invariants predicted by a horseshoe template, those calculated from the data sets shown in Figs. 2, and data from the model constructed in Section III. Some of these periodic orbits are shown in Fig. 6. These results provide good evidence that—to within the resolution of the measurements and in the parameter regime considered—the orbits of the string and the model are governed by a horseshoe template with zero global torsion.

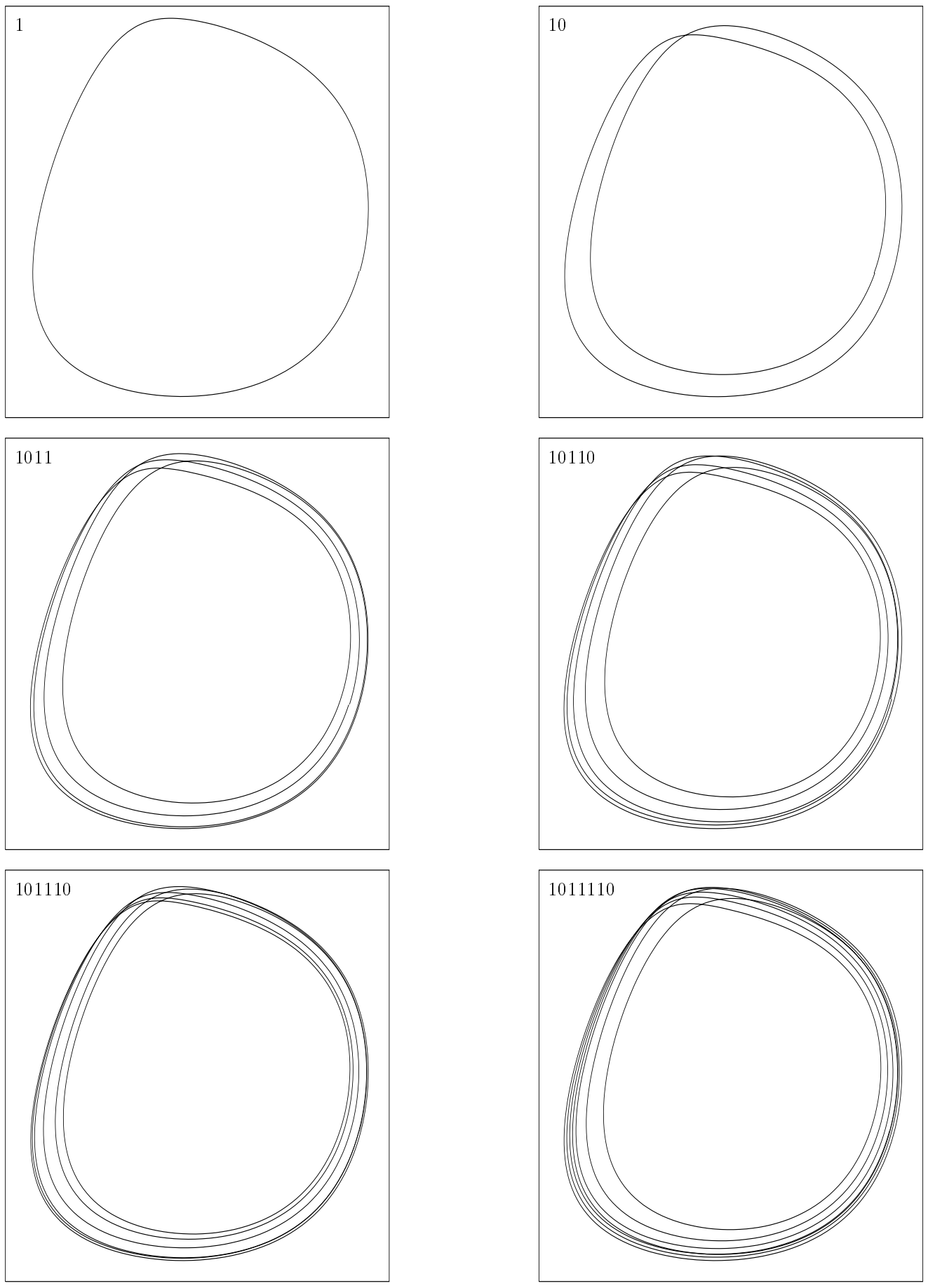

FIG. 6. Some periodic orbits extracted from a chaotic time series from the string experiment.

It is encouraging that the model agrees in this first topological check. Evidently, the analytic properties of the model class, along with the initial conditions provided by the experimental data, are sufficient to determine (at least approximately) the fixed points which are in the vicinity of the sampled flow.

### B. Symbol Plane

To get a topological "road map" of the data we use an empirical procedure to construct the symbol plane generated by a single chaotic trajectory [15]. We construct two symbol planes, one for data from the experiment (Fig. 2(b)), consisting of 725 points in the return map, and one for the model, consisting of 4500 points in the return map. To construct the symbol plane we first convert the chaotic trajectory to a symbolic string of 0's and 1's, depending on whether the orbit passes to the left or right of the maximum point of the first return map shown in the inset of Fig. 2(b). As shown in the next Subsection, this empirical symbolic partition is fine enough to distinguish orbits at least up to period eleven. Now, since we know the topological form (the template) of the chaotic set, we can use kneading theory to determine the "well ordered" [15,42] symbol sequences needed to construct the symbolic plane.

After conversion the scalar data set has become a symbol string of the form

$$\mathbf{s} = \ldots s_{-3}s_{-2}s_{-1}s_0.s_1s_2s_3\ldots$$

where symbols to the left and right of $s_0$ are the past and future symbols respectively. The coordinates of the symbol plane are calculated from the well ordered past ($c_i$) and future ($b_i$) symbols as follows:

$$x(\mathbf{s}) = \sum_{i=1}^{D} \frac{b_i}{2^i}, \qquad b_i = \sum_{j=1}^{i} s_j \bmod 2$$

and

$$y(\mathbf{s}) = \sum_{i=0}^{D-1} \frac{c_i}{2^i}, \qquad c_i = \sum_{j=0}^{i+1} s_{-j} \bmod 2.$$

If $\mathbf{s}$ is an infinite symbol string generated by a chaotic orbit, then $D$ is infinity in the above sums. However, since we are dealing with finite data sets, we approximate the symbol plane coordinates of a point by taking $D = 16$. In this way we can use a finite symbol string from a chaotic trajectory to generate a sequence of points on the symbol plane.

The resulting plots for both the experimental and model data are shown in Fig. 7. These plots suggest several nontrivial observations about the topological properties of the flows producing these data sets. First, it appears that a good approximation to the "pruning front" [15] is a single vertical line (a schematic for this pruning front is illustrated in Fig. 7), at least to within the resolution of our experimental measurements and data processing procedures. This indicates that, at the topological level, the symbolic dynamics of the system are well-described by unimodal map theory, with the pruning determined by a single topological parameter, the kneading invariant [15,42], which we can attempt to approximate [43]. Second, the close similarity between the plot generated by experimental data, and model data, again provides striking evidence that the empirical model correctly captures the topological properties of the flow. Moreover, we can estimate the closeness of this topological fit by comparing estimates of the kneading invariant from the experimental data and model data. Third, the simple form of these plots, and the conjectured (approximate) pruning front, provides yet another powerful self-consistency check that the template (and hence the formula for well-ordering the symbol sequences) is identified correctly.

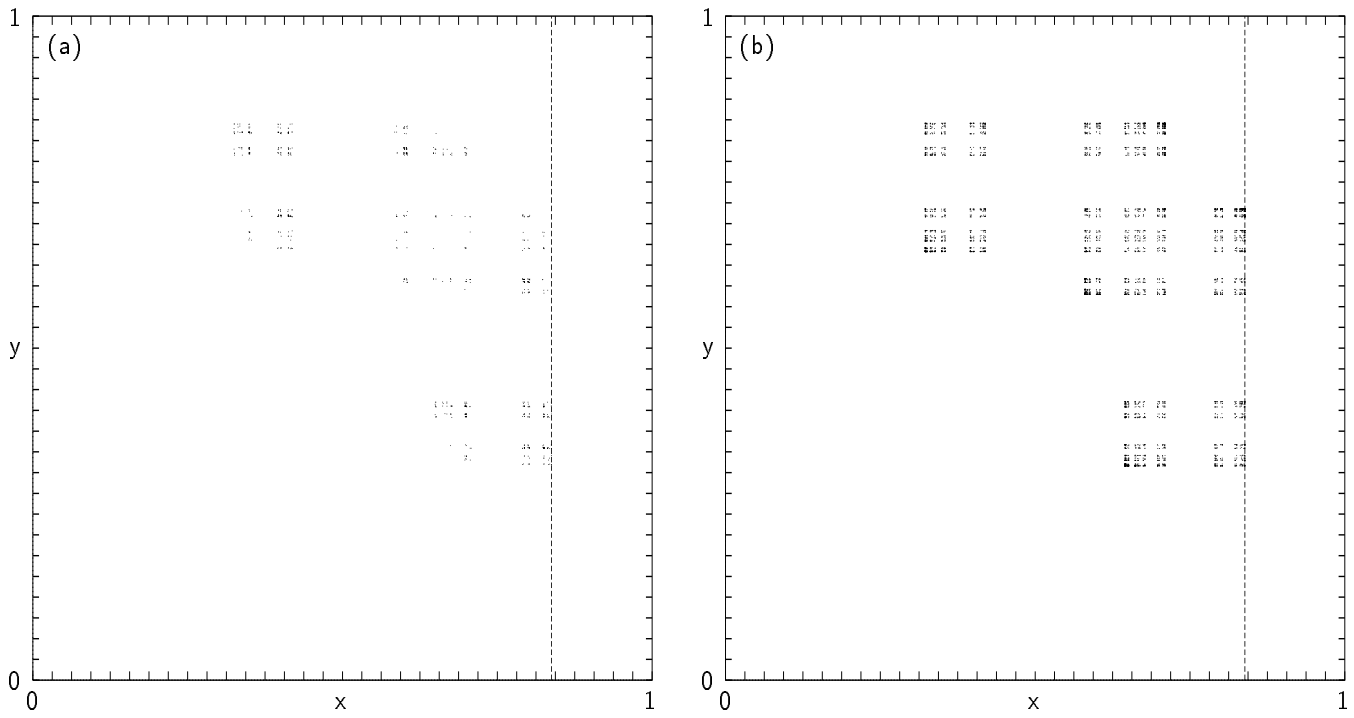

FIG. 7. Symbol planes generated by chaotic time series from a string: (a) Experimental data (b); Model (synthetic) data.

In the next section we extract periodic orbits from the chaotic time series and show how they can be used in this example to systematically approximate the vertical pruning front suggested by Fig. 7.

### C. Periodic orbit spectra

To extract the (approximate) periodic orbits by the method of close recurrence we first convert the next impact map from the coordinate values directly into a symbol sequence. In this particular instance, we found that an adequate symbolic description (at least up to period eleven orbits, or approximately one part in $2^{11}$) is obtained by choosing the maximum value of the next impact shown in the insets of in Fig. 2. Orbits passing to the left of the maximum are labeled '0', and those to the right are labeled '1'. Next we search this symbolic encoding for each and every periodic symbol string. Every time a periodic symbol string is found we calculate its (normalized) recurrence and then save the instance of the orbit with the best recurrence. The advantage of this procedure of periodic orbit extraction is that it is exhaustive. We search for every possible orbit up to a given period. In these studies we searched for all orbits of period one through eleven.

The resulting spectrum of periodic orbits for both experimental and model data sets is shown in Table 1. The orbits which are present in (the full shift) complete hyperbolic system, and not present in the Table 1, are said to be *pruned*. Two topological invariants (the linking number and the exponent sum of positive braids) of the extracted orbits are calculated and compared with those of a horseshoe (see Table 1). There are no discrepancies on the orbits in which the crossings can be unambiguously resolved. This indicates that, at least to this level of resolution, the template is correctly identified and the symbolic partition is adequate. Moreover, the template for the string experiment remains unchanged for the two different experimental parameter settings examined, though the pruning is much more severe in data set (a) than it is in data set (c). The symbolic label (up to braid type) can also be determined—and used as yet another self-consistency check—by considering a simple and easily computable braid invariants. As pointed out by Hall [41], the exponent sum (simply the sum of braid crossings in our example) is a complete invariant for horseshoe braids up to period eight. Also, an inspection of the exponent sums as a function of period reveals that the exponent sum manages to distinguish most of the pseudo-Anosov horseshoe braids of periods nine, ten, and eleven as well [43]. The symbolics determined by the braid type, and that determined by the empirical partition are consistent for all the orbits listed in Table 1. Our goal now is to predict as best as possible the pruned spectrum from the chaotic time series.

TABLE I. Spectrum of low period orbits extracted from chaotic time series (all orbits with $\epsilon < 0.03$) are shown). Extracted orbits, their exponent sum, one-dimensional topological entropy, and their (best) normalized recurrence are recorded. Data set (a) (experimental with 500 points in return map); Data set (b) is the dynamically cleaned version of data set (c); Data set (c) (experimental with 725 points in return map); Data set (d) is from the empirical model of data set (c)—the "synthetic data" (4500 points in return map).

| P | $c_P$ | es | $h_1$ | (a) | (b) | (c) | (d) |
|---|---|---|---|---|---|---|---|
| $s_1^1$ | 1 | 0 | 0 | – | 0.02192 | 0.01383 | 0.01880 |
| $s_2^1$ | 10 | 1 | 0 | 0.01796 | 0.01459 | 0.01773 | 0.02929 |
| $s_4^1$ | 1011 | 5 | 0 | 0.00840 | 0.01287 | 0.00993 | 0.02386 |
| $s_5^1$ | 10110 | 8 | 0.414 | – | – | – | 0.02678 |
| $s_5^1$ | 10111 | 8 | 0.414 | – | – | – | 0.00863 |
| $s_6^1$ | 101110 | 13 | 0.241 | 0.00780 | 0.00433 | 0.00849 | 0.01106 |
| $s_6^1$ | 101111 | 13 | 0.241 | 0.00675 | 0.00381 | 0.01350 | 0.01202 |
| $s_7^1$ | 1011110 | 18 | 0.382 | – | 0.00634 | 0.00917 | 0.00994 |
| $s_7^1$ | 1011111 | 18 | 0.382 | – | – | – | 0.01260 |
| $s_8^1$ | 10111010 | 23 | 0 | 0.00750 | 0.00299 | 0.00789 | 0.01903 |
| $s_8^2$ | 10111110 | 25 | 0.305 | – | 0.02241 | 0.01770 | 0.02087 |
| $s_8^2$ | 10111111 | 25 | 0.305 | – | – | – | 0.01264 |
| $s_9^1$ | 101111110 | 32 | 0.366 | – | 0.00893 | 0.01421 | 0.00662 |
| $s_9^1$ | 101111111 | 32 | 0.366 | – | – | – | 0.02458 |
| $s_9^2$ | 101111010 | 30 | 0.397 | – | 0.01864 | 0.02638 | 0.02201 |
| $s_9^2$ | 101111011 | 30 | 0.397 | – | – | – | 0.02336 |
| $s_{10}^1$ | 1011101010 | 37 | 0.207 | 0.00752 | 0.00225 | 0.00635 | 0.01260 |
| $s_{10}^1$ | 1011101011 | 37 | 0.207 | 0.00511 | 0.01560 | 0.01826 | 0.02052 |
| $s_{10}^2$ | 1011111010 | 39 | 0.272 | – | 0.01694 | 0.00859 | 0.00452 |
| $s_{10}^2$ | 1011111011 | 39 | 0.272 | 0.02766 | 0.01312 | 0.01875 | 0.01438 |
| $s_{10}^3$ | 1011111110 | 41 | 0.328 | – | 0.00622 | 0.02037 | 0.01512 |
| $s_{10}^3$ | 1011111111 | 41 | 0.328 | – | – | – | 0.00772 |
| $s_{11}^1$ | 10111111110 | 50 | 0.357 | – | 0.02365 | 0.01339 | 0.02657 |
| $s_{11}^1$ | 10111111111 | 50 | 0.357 | – | – | – | 0.00974 |
| $s_{11}^2$ | 10111111010 | 48 | 0.374 | – | – | – | 0.00960 |
| $s_{11}^2$ | 10111111011 | 48 | 0.374 | – | – | – | 0.01491 |
| $s_{11}^3$ | 10111101010 | 46 | 0.390 | – | – | 0.02305 | 0.01199 |
| $s_{11}^3$ | 10111101011 | 46 | 0.390 | – | – | – | 0.01811 |
| $s_{11}^4$ | 10111101110 | 46 | 0.403 | – | – | 0.01483 | 0.01780 |
| $s_{11}^4$ | 10111101111 | 46 | 0.403 | – | – | – | 0.01600 |

Before we discuss pruning, though, it is useful to consider the number of distinct periodic orbits extracted as a function of the number of points in the return map. This is shown for the model data in Table 2. As expected, the number of orbits that can be extracted increases with the number of points in the return map. More importantly, Table 2 strongly suggests that using the method of close recurrence it is possible to obtain all the low period orbits embedded within the strange attractor. For instance, after $10^5$ points are examined, we see that no new periodic orbits are found below period seven. Similarly, after examining $5 \times 10^5$ points we believe we have found all orbits up to period nine. These results also caution us when working with smaller samples such as those in data set (a) with 500 points in the return map, and data set (b) with 725 points in the return map—the extracted orbit spectrum is expected to miss orbits either because the orbit is pruned (it is not in the strange set) or because the sample of the strange set we are examining fails to provide a close enough coverage over the whole attractor.

As suggested by the symbol plane diagram (Fig. 7), unimodal map theory should be sufficient to explain these periodic orbit spectra. To test this hypothesis, we first examine the periodic orbit spectra from the model data set and locate the maximal (rightmost) periodic point in terms of unimodal theory. It is a period five orbit with symbolic name 10110, and indicated by the label (up to saddle-node pairs) of $s_5^1$. In unimodal theory, the $s_5^1$ orbit *forces* [44] the orbits $s_{11}^4$, $s_9^2$, $s_{11}^3$, $s_7^1$, $s_{11}^2$, $s_9^1$, $s_{11}^1$, $s_{10}^3$, $s_8^2$, $s_{10}^2$, $s_6^1$, $s_{10}^1$, $s_8^1$, $s_4^1$, $s_2^1$, and $s_1^1$. The notation identifies (up to saddle-node pairs) the period of each orbit in the subscript, and the value in terms of unimodal ordering (within each period) in the superscript. This theoretically determined spectrum agrees with that found in Table 1. In fact, both partners of all the saddle-node pairs are present except for the period one orbit pair. In this case, though, it is known that the period one orbit with symbolic label '0' is not present in the strange attractor, though it is present in the flow.

TABLE II. Number of distinct periodic orbits of a given period extracted as a function of the sample size ($\times\ 10^3$).

| P | 5 | 10 | 50 | 100 | 500 | 816 |
|---|---|---|---|---|---|---|
| 1 | 0 | 0 | 1 | 1 | 1 | 1 |
| 2 | 0 | 0 | 0 | 0 | 1 | 1 |
| 3 | 0 | 0 | 0 | 0 | 0 | 0 |
| 4 | 0 | 1 | 1 | 1 | 1 | 1 |
| 5 | 1 | 1 | 1 | 2 | 2 | 2 |
| 6 | 0 | 0 | 2 | 2 | 2 | 2 |
| 7 | 0 | 1 | 2 | 2 | 2 | 2 |
| 8 | 0 | 1 | 1 | 2 | 3 | 3 |
| 9 | 0 | 0 | 1 | 3 | 4 | 4 |
| 10 | 0 | 1 | 2 | 4 | 6 | 6 |
| 11 | 2 | 2 | 2 | 4 | 8 | 8 |

Moreover, we can now use the maximal periodic orbit to estimate the itinerary of the kneading sequence in the model data set, $\kappa_m \approx \overline{10110}$, or in two dimensions by the (vertical) pruning front specified by $\overline{0}_1^0.\overline{10110}$. The horizontal coordinate of the (maximal) point of this periodic orbit on the symbol plane is $x(\mathbf{s}) \approx 0.8485$. An estimate of the kneading invariant also provides an estimate of the (one-dimensional) topological entropy [45,46]: $h_1(s_5^1) \approx 0.4140$.

A similar analysis of the experimental data set (c) shows that the maximal periodic orbit in the unimodal ordering is $s_{11}^4$ which forces $s_9^2$, $s_{11}^3$, $s_7^1$, $s_{11}^2$, $s_9^1$, $s_{11}^1$, $s_{10}^3$, $s_8^2$, $s_{10}^2$, $s_6^1$, $s_{10}^1$, $s_8^1$, $s_4^1$, $s_2^1$, and $s_1^1$. Up to period six, all the orbits forced by $s_{11}^4$ are present. Nine orbits are missing between periods seven through eleven. But we believe that these higher period orbits are missing due to the small sample size. We use the maximal orbit to approximate the itinerary of the kneading sequence in the experimental data set as $\kappa_b \approx \overline{10111101111}$, or in two dimensions by the (vertical) pruning front specified by $\overline{0}_1^0.\overline{10111101111}$. The horizontal coordinate of the maximal point of this periodic orbit on the symbol plane is $x(\mathbf{s}) \approx 0.8385$. An estimate of the one-dimensional topological entropy derived from the experimental data is: $h_1(s_{11}^4) \approx 0.4032$.

Time series from the model and experiment are close in both their topological form and topological parameter value(s). The model, though, does appear to determine parameter values which over estimate the topological entropy. This difference in entropy could be due to several sources. First, uncertainties due to noise and parametric drift in the experimental data set place inherent limits on the modeling procedure's accuracy. Second, low-period orbits, perhaps of higher entropy than those extracted, may be present, but are not found because of the limited sample size of the experimental data. Third, the estimation in the entropy can not be any finer than that allowed by the order of the periodic orbit approximation. As an examination of Table I indicates, the difference in entropy between the two periodic orbits considered (up to period eleven), which are close in the unimodal ordering, is roughly 0.01. This last point suggests that the estimated difference in entropy between the experimental data and model data is an upper bound on the difference. The entropy between the two process may, in fact, be closer than this estimate indicates.

## V. CONCLUDING REMARKS

In this paper we create an empirical global model of a data set from a string experiment and show that the model captures the dynamics implicit within the data set in at least two significant respects: the model and data can be made to synchronize [19,26], and they share quantifiable common topological properties. We also suggest that it is a sensible and desirable procedure to use synthetic data from the model to characterize the system in ways which may not be readily accessible from the data set alone. In particular we discuss how to characterize the system at the topological level by its periodic orbit spectrum, and at the metric level by its Lyapunov spectrum. More generally, we illustrate how to characterize a chaotic invariant set both by its general form (a template) and specific topological parameter values (kneading sequences) estimated from the spectrum of periodic orbits of the chaotic attractor, or from an empirically constructed "pruning front" [15].

Global, empirically constructed, analytic models open the door to a host of practical applications in systems analysis, identification, and control [19]. In particular, we emphasize how synthetic data sets permit the calculation of quantities requiring data with a larger sample size, or lower noise, than may be directly available from the experimental data alone [34]. In addition, we would also like to emphasize how a synchronized model of a system can be coupled directly to a data stream permitting a kind of model-based real-time dynamic filter [47].

## ACKNOWLEDGMENTS

N. B. T. thanks David Griffiths for drawing his attention to "The Bean-Field," Kevin Judd and Alistair Mess for talks about MDL and modeling, and Bob Gilmore for assistance in calculating the entropies. P. W.'s work was supported by a Computational Science Graduate Fellowship from the Department of Energy.

---